%% file: kashiwagi_encoderprompting.tex
\documentclass{Interspeech2024}

\usepackage{cite}
\usepackage{amssymb}
\usepackage{color}
\usepackage{comment}

\newcommand{\argmax}{\mathop{\rm arg~max}\limits}

\interspeechcameraready

\title{Rapid Language Adaptation for Multilingual E2E Speech Recognition Using Encoder Prompting}

\name[affiliation={1}]{Yosuke}{Kashiwagi}
\name[affiliation={1}]{Hayato}{Futami}
\name[affiliation={1}]{Emiru}{Tsunoo}
\name[affiliation={2}]{Siddhant}{Arora}
\name[affiliation={2}]{Shinji}{Watanabe}

\address{
  $^1$Sony Group Corporation, Japan, $^2$Carnegie Mellon University, USA}
\email{yosuke.kashiwagi@sony.com}


\begin{document}

\maketitle

\begin{abstract}

    End-to-end multilingual speech recognition models handle multiple languages through a single model, often incorporating language identification to automatically detect the language of incoming speech.
    Since the common scenario is where the language is already known, these models can perform as language-specific by using language information as prompts, which is particularly beneficial for attention-based encoder-decoder architectures.
    However, the Connectionist Temporal Classification (CTC) approach, which enhances recognition via joint decoding and multi-task training, does not normally incorporate language prompts due to its conditionally independent output tokens.
    To overcome this, we introduce an encoder prompting technique within the self-conditioned CTC framework, enabling language-specific adaptation of the CTC model in a zero-shot manner. Our method has shown to significantly reduce errors by 28\% on average and by 41\% on low-resource languages.
    
\end{abstract}

\section{Introduction}
In recent years, multilingual speech recognition models have emerged, taking advantage of massive computational resources and large amounts of data \cite{watanabe2017language,toshniwal2018multilingual,radford2023robust,peng2023reproducing,peng2024owsm,zhang2023google,tjandra2023massively,pratap2023scaling}.
These are efficiently trained by aggregating large amounts of data into a single model.
For example, Whisper supports more than 100 languages in a single model \cite{radford2023robust}.
Open Whisper-style speech model is a model that reproduces a Whisper-like model with open data \cite{peng2023reproducing,peng2024owsm}.
Google universal speech model uses large amounts of unpaired data to improve recognition performance for low-resource languages \cite{zhang2023google}.
Meta has proposed MMS and is attempting to extend it to over 1000 languages \cite{tjandra2023massively,pratap2023scaling}.
Multilingual speech recognition has been reported to be advantageous in terms of data efficiency, especially in low-resource languages \cite{dalmia2018sequence,zhou2018multilingual}.

Language identification is often provided as a secondary function of multilingual speech recognition.
In addition to simply identify the language of speech, there are also attempts to detect language switches in multi-speaker conversations \cite{seki2019end}.
However, language identification is not always required for daily use in speech recognition systems.
When individuals use speech recognition, the language they speak is often predetermined.
E2E multilingual speech recognition provides a function to adapt the model for a specific language by giving the target language ID as a prompt to the decoder, especially for the attention-based encoder decoder \cite{radford2023robust}.
It has been reported that providing a language ID can significantly improve multilingual recognition performance \cite{9003870,li2018multi,yang2023adapting,li2019towards,winata2020meta,dalmia2021transformer}.

On the other hand, it is also reported that multi-task training and joint-decoding using Connectionist Temporal Classification (CTC) can improve recognition performance in E2E multi-language speech recognition \cite{graves2006connectionist,hori2017joint,peng2023reproducing}.
In this joint-decoding framework, we hypothesize that the performance can be further improved by providing language IDs while computing the CTC output.
However, since the CTC output at each frame is conditionally independent, it is not possible to adapt the recognition results by providing a language ID as the decoder.

Self-conditioned CTC (SC-CTC) is proposed to mitigate the conditional independence of CTC.
SC-CTC calculates CTC loss in the middle layer of the encoder and adds the intermediate prediction to the input of the next encoder layer \cite{nozaki2021relaxing}.
Many variants of SC-CTC have been proposed to improve its performance \cite{komatsu2022better,nakagome2022interaug,li2023enhancing}.
Related to multilingual speech recognition, hierarchically changing the importance of language tokens during training has also been proposed \cite{chen2023improving}.
However, this previous study targets training methods and not language adaptation during inference.

Our proposed method uses the general SC-CTC framework during training and can be adapted quickly by simply providing prompts during inference.
Prompting is accomplished by modifying the probabilities of language IDs of the token sequence estimated in the intermediate layer.
Prior work adapts by adding linguistic information to the input \cite{li2023enhancing}.
However, the prior work requires a significant change in structure during training and does not assume the case where no linguistic information is given.
On the other hand, our method can operate as a general self-conditioned model if no prompt is given during inference.
We confirmed the effectiveness of our proposed method with the Common Voice, VoxForge and FLEURS corpus \cite{ardila2020common,voxforge,conneau2023fleurs}.
Experiments showed that our method achieved an average relative error reduction of 28\% on Common Voice data.
It was also particularly effective for extremely low-resource languages, achieving an error reduction of 41\% for languages with less than 5 hours of training data.

\section{Related works}

\subsection{E2E multilingual ASR}

\begin{figure}[t]
  \centering
  \includegraphics[width=\linewidth]{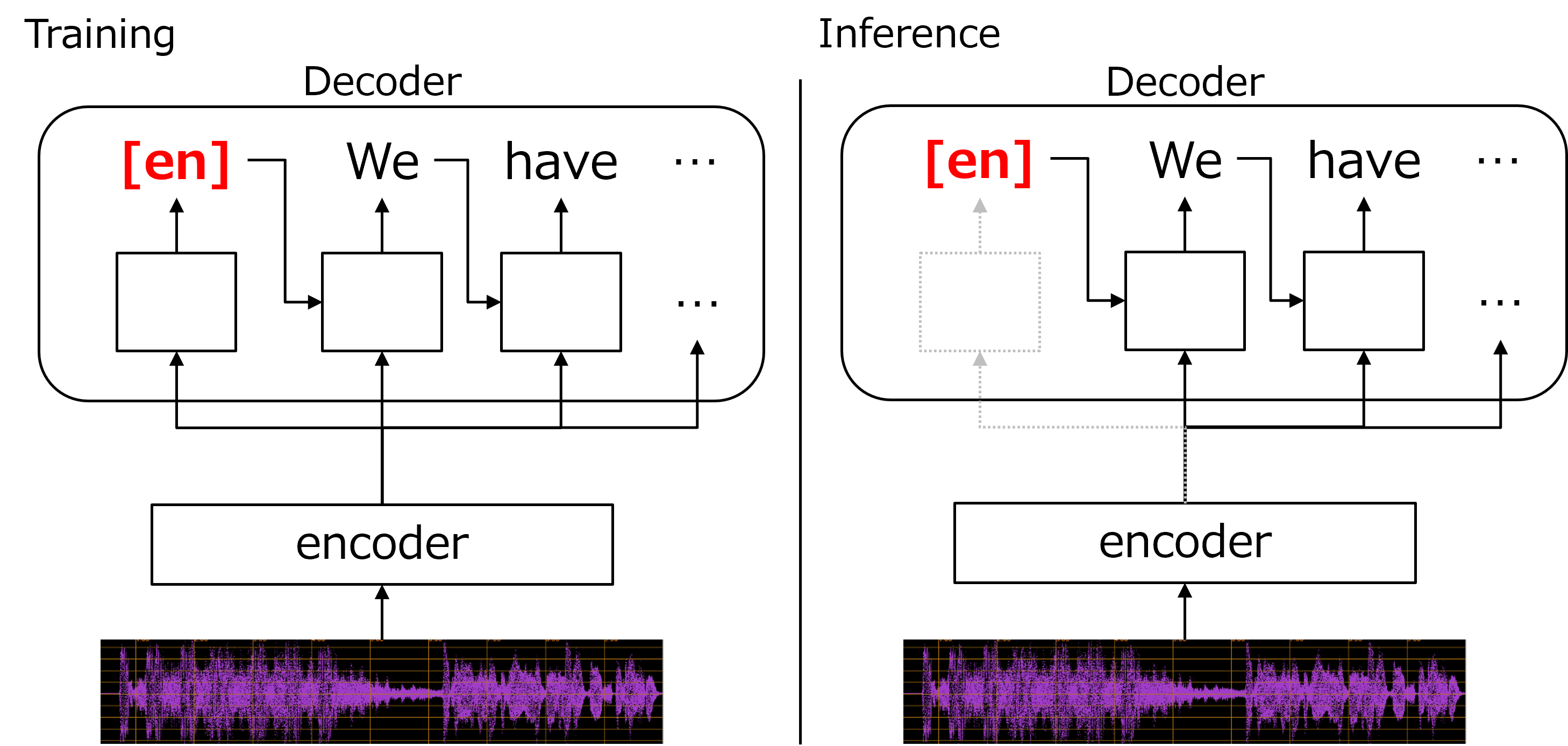}
  \caption{Multilingual ASR with language prompts added to the decoder.}
  \label{fig:multilingualASR}
\end{figure}

Multilingual speech recognition has received a great deal of attention in recent years, and many studies and models have been published \cite{watanabe2017language,toshniwal2018multilingual,radford2023robust,peng2023reproducing,peng2024owsm,zhang2023google,tjandra2023massively,pratap2023scaling}.
By training many languages with a single model, the recognition performance is improved, especially for languages with a small number of data \cite{dalmia2018sequence,zhou2018multilingual}.
In our study, we further focus on the E2E multilingual model, which estimates language IDs simultaneously with the text.
As shown in Figure \ref{fig:multilingualASR}, this model uses the token sequence with the language ID added to the beginning of the transcription in training.
During inference, users can choose to explicitly specify the language or not.
If they want to specify the language, they can fix the language ID; if not, they can use the model to infer the language ID as well.
This allows the model to be flexibly controlled according to the situation in which it is used.
For example, in situations where the input speech comes from an unspecified number of speakers, such as on the internet, supporting as many languages as possible will provide many people with the opportunity to interact freely.
On the other hand, in situations where the language of input speech is predetermined, such as on a personal device, the recognition accuracy can be improved by using the system specialized for a particular language.

\subsection{Self-conditioned CTC}

CTC is a non-autoregressive model in which each frame is conditionally independent \cite{graves2006connectionist}.
This is an advantage of fast computation, but it can also be a disadvantage.
Since conditional independence is a strong assumption, generating text sequences using CTC alone can be suboptimal.
Therefore, CTC is often used in combination with an attention decoder \cite{hori2017joint}.
It has been reported that multitask training and joint-decoding with CTC are effective for E2E speech recognition.
This technique is useful not only for improving recognition accuracy but also for contributing to training stability.
On the other hand, SC-CTC has been proposed to mitigate the conditional independence assumption \cite{nozaki2021relaxing}.
SC-CTC is an extension of intermediate CTC (InterCTC) \cite{lee2021intermediate}.
InterCTC is a method of performing multitask training by calculating CTC losses in the intermediate layer of the encoder.
This has been shown to stabilize encoder training.
SC-CTC further adds the estimated CTC labels at the intermediate layer to the next layer through a linear transformation.
It has been reported that this subnetwork improves the recognition performance of the CTC model.
Furthermore, SC-CTC improves the joint-decoding performance using CTC and an attention-based encoder-decoder (AED) for multilingual ASR as shown in Figure \ref{fig:selfconditionedctc} \cite{chen2023improving}.

\section{Proposed method}
\subsection{Rapid language adaptation}
\begin{figure}[t]
  \centering
  \includegraphics[width=\linewidth]{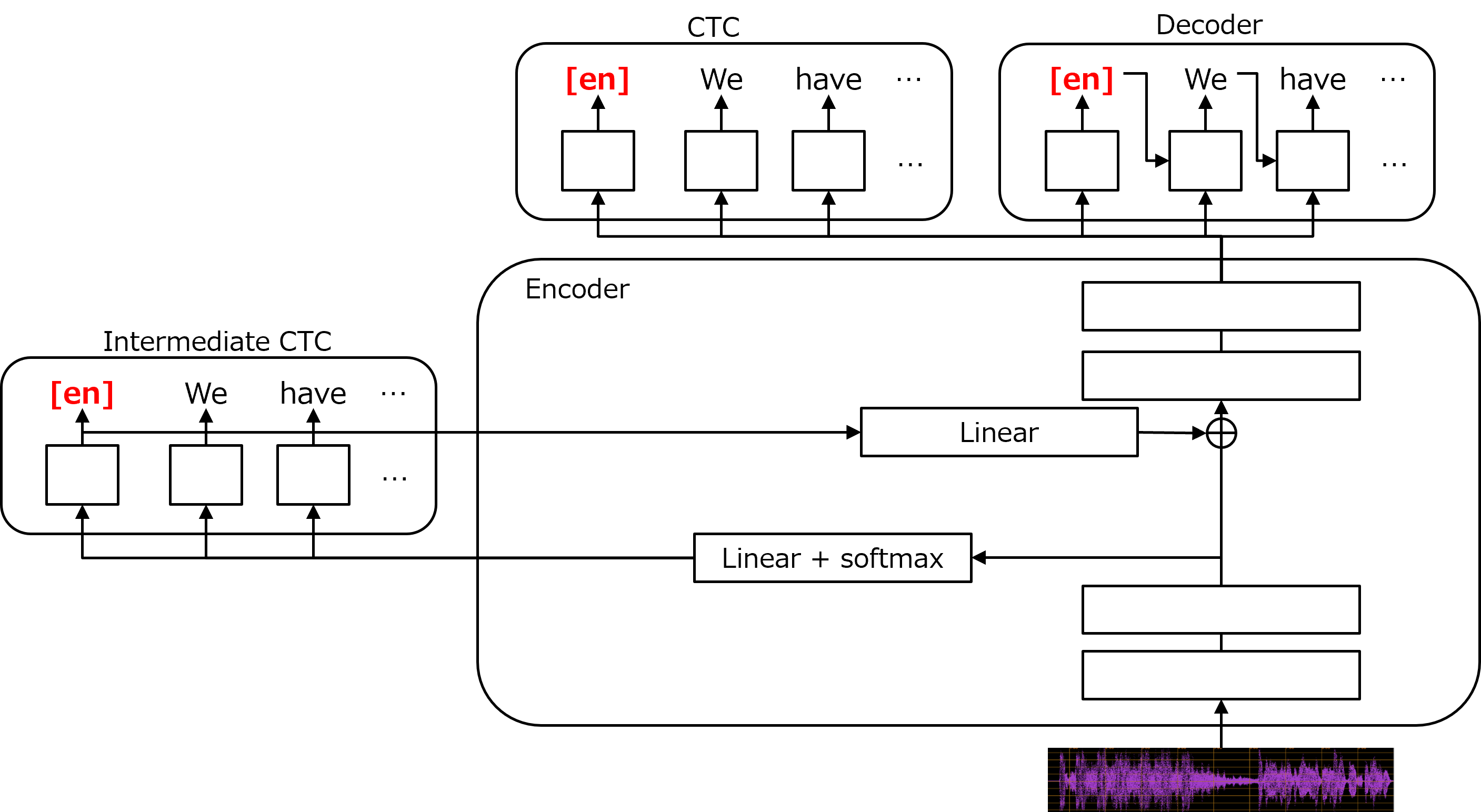}
  \caption{Joint CTC-AED model with self-conditioned CTC for multilingual ASR.}
  \label{fig:selfconditionedctc}
\end{figure}
\begin{figure}[t]
  \centering
  \includegraphics[width=\linewidth]{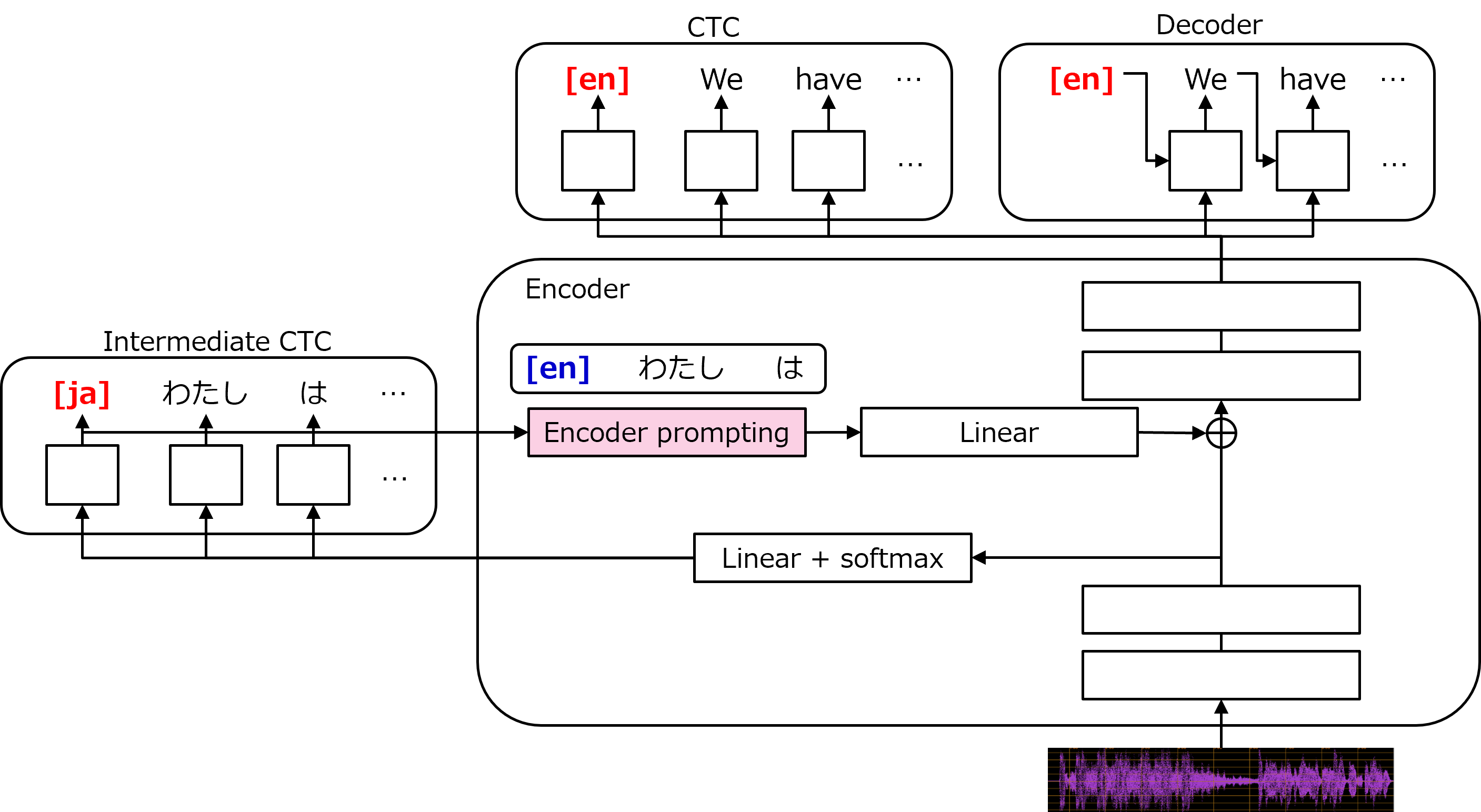}
  \caption{Proposed rapid language adaptation with encoder prompting during inference.}
  \label{fig:rapidadaptation}
\end{figure}

\begin{table*}[thbp]
  \caption{Comparison of related baseline models and our proposed encoder prompting (Sec. 3.2) on Common Voice evaluation set in CER(\%). The languages are divided into high, middle, low, and extreme low according to the amount of data. High is for languages with more than 100 hours of data, middle is for 20-100 hours, low is for 5-20 hours, and extremely low is for less than 5 hours.}
  \label{table:results1}
  \centering
  \begin{tabular}{l|l|ccc|cccc|c}
    \hline
        ID & encoder type & \begin{tabular}{c}
joint \\
decoding
\end{tabular}  & \begin{tabular}{c}
decoder \\
prompting
\end{tabular} & \begin{tabular}{c}
encoder \\
prompting
\end{tabular} & high & middle & low & ex.low & avg. \\
    \hline
    (a) & Transformer &    &    &    &   \hphantom{0}6.1 & 10.5 & 26.7 & 39.7 & 22.9
 \\
    (b) & Transformer &  $\checkmark$ &    &    & \hphantom{0}6.2 & 10.3 & 26.2 & 39.6 & 22.6 \\
    (c) & InterCTC & $\checkmark$  &    &    & \hphantom{0}6.1 & 11.4 & 28.1 & 40.6 & 23.7 \\
    (d) & SC-CTC &  $\checkmark$  &    &    &    \hphantom{0}6.1 & 11.2 & 27.9 & 42.5 & 24.2 \\
    \hline
    (e) & Transformer &    &  $\checkmark$  &    & \hphantom{0}5.9 & \hphantom{0}9.3 & 21.4 & 35.5 & 19.9 \\
    (f) & Transformer &  $\checkmark$  &   $\checkmark$ &    &  \hphantom{0}6.1 & \hphantom{0}9.7 & 23.5 & 35.0  & 20.4 \\
    (g) & InterCTC &  $\checkmark$  &  $\checkmark$  &    & \hphantom{0}6.0 & 10.5 & 25.5 & 35.9 & 21.3\\
    (h) & SC-CTC &  $\checkmark$  &  $\checkmark$  &    &  \hphantom{0}6.1 & 10.4 & 26.2 & 38.7 & 22.4 \\
    \hline
    (i) & SC-CTC &  $\checkmark$  &  $\checkmark$  &  \textit{Replacement} &  \bf{\hphantom{0}5.8} & \bf{\hphantom{0}8.7} & \bf{18.1} & \bf{20.9} & \bf{14.3} \\
    (j) & SC-CTC &  $\checkmark$  &  $\checkmark$  &  \textit{Aggregation} & \bf{\hphantom{0}5.8} & \bf{\hphantom{0}8.7} & \bf{18.1} & 21.0 & \bf{14.3} \\
    (k) & SC-CTC &  $\checkmark$  &  $\checkmark$  &  \textit{Prefix}  & \hphantom{0}6.1 & 10.4 & 26.2 & 38.7 & 22.4 \\
    \hline
  \end{tabular}
\end{table*}

\begin{table}[t!]
  \caption{Language ID accuracy (\%) for various encoders. The ID of each row are unified with Table \ref{table:results1}.}
  \label{table:results2}
  \centering
  \begin{tabular}{l|l|cccc}
    \hline
         ID & encoder type & high & middle & low & ex.low\\
    \hline
     (b) & Transformer & 98.2 & 93.0 & 78.5 & 43.6

\\
     (c) & InterCTC & 97.7 & 89.9 & 73.8 & 38.4 \\
     (d) & SC-CTC &  97.7 & 90.4 & 73.6 & 33.9 \\
    \hline
  \end{tabular}
\end{table}

There are two ways to use E2E multilingual speech recognition.
One is to perform language identification and speech recognition simultaneously, which is targeted for environments where speech from an unspecified number of speakers is input.
The other is to use the model with a predefined language, which is intended for use on personal devices or within a limited community scenario where the language of the speaker is predetermined.
We propose encoder prompting, a new rapid language adaptation method targeting the latter case where the input language is predetermined.
The encoder prompting is to control not only the decoder but also the encoder in a prompt-based manner.
To achieve this, a SC-CTC-based joint ASR CTC model is first trained, as shown in Figure \ref{fig:selfconditionedctc}.
During inference, language IDs are modified to the predefined target language in the intermediate layer of the SC-CTC, as shown in Figure \ref{fig:rapidadaptation} (Encoder Prompting).
Sec. \ref{sec:encoder_prompting} describes how to modify the labels.
After that, the modified token sequence is returned to the middle layer through a linear transformation similar to SC-CTC.
In Figure \ref{fig:rapidadaptation}, the input speech is English, but the intermediate layer has the highest probability of being Japanese.
In this case, the model knows the input language is English because it is predetermined by the user.
Therefore, the output of the intermediate layer is modified so that the probability of an English token is the highest.
Since the proposed method requires modification only during inference, it can be used to quickly adapt an already trained model.
Furthermore, the modified token sequence affects the output of the encoder, so it is reflected in the output of both the attention decoder as well as the CTC.

\subsection{Encoder prompting}
\label{sec:encoder_prompting}
We investigated three ways to modify the output of CTC to reflect the language information.
The first is to modify only the frames with the highest language ID probability as:
\begin{align}
\hat{p}_t(k) = \begin{cases}
\text{OneHot}(k_{\text{target}}) & \argmax_{k'} p_t(k') \in K_{\text{LID}} \\
p_t(k) & \text{otherwise}
\end{cases},
\label{eq:replacement}
\end{align}
where $p_t(k)$ is the intermediate layer's output probability for token $k$ in time $t$ and $\hat{p}_t(k)$ is the modified probability after encoder prompting.
$K_{\text{LID}}$ is the set of tokens corresponding to the language ID and $k_{\text{target}}$ is the target language ID.
$\text{OneHot}(k_{\text{target}})$ describes the Kronecker delta.
If $k=k_{\text{target}}$, it will be $1$; otherwise it will be $0$.
In this approach, the CTC output frames representing the language information are overwritten by the appropriate target language using a one-hot vector.
We call this approach as \textit{Replacement}.
Although this approach requires minimal modification, it does not modify the linguistic information that would have remained in other frames.
This is because the probability of language IDs is maximal only at the beginning of an utterance.
However, language information that remains in other frames, even with low probability, may affect recognition performance.

Therefore, we further propose \textit{Aggregation} approach that modifies Eq.(\ref{eq:replacement}) to aggregate the probabilities of the other language IDs to the appropriate language ID in all frames as:
\begin{align}
\hat{p}_t(k) = \begin{cases}
p_t(k) & k \notin K_{\text{LID}} \\
\sum_{k' \in K_{\text{LID}}} p_t(k') & k=k_{\text{target}} \\
0 & \text{otherwise}
\end{cases}.
\label{eq:aggregation}
\end{align}
This approach can correct the language information in all frames.
Since the same process is applied to all frames, the implementation is simple.
However, \textit{Aggregation} may result in the probability of language IDs being too high in unintended frames.
If this were the case, it would be an unnatural label for multilingual speech recognition.

\begin{table}[t!]
  \caption{Investigation of using soft prompting (Sec. 3.3) on Common Voice data in CER(\%). In all conditions, encoder is SC-CTC and joint-decoding is used. The soft prompting is applied to 3 languages during inference (English, Chinese and Japanese). The ID of each row are unified with Table \ref{table:results1}.}
  \label{table:results3}
  \centering
  \begin{tabular}{l|cc|ccc}
    \hline
        ID & decoder & encoder & & &  \\
        & prompting & prompting & EN & CN & JA \\
    \hline
    (h) &  $\checkmark$  &   & 8.5 & 23.4 & 48.2 \\
    (j) &  $\checkmark$  &  \textit{Aggregation}  &  8.4 & 22.6  & 36.5  \\
    \hline
    (d) &  &    &  8.5 & 25.0 & 51.2 \\
    (l) &    &  \textit{Soft aggregation}  & \bf{8.4}  & \bf{23.2} & \bf{43.4} \\
    \hline
  \end{tabular}
\end{table}

On the other hand, unlike decoder prompts, \textit{Replacement} and \textit{Aggregation} do not allow the input of natural language prompts of any length.
To mitigate this, as a third approach, we propose \textit{Prefix} method that overwrites only the minimum number of frames necessary to represent the prompt from the beginning of the utterance.
In our experiments, language ID requires only one frame.
Therefore, we modify Eq.(\ref{eq:replacement}) such that only the head frame is replaced as:
\begin{align}
\hat{p}_{t=0}(k) = \text{OneHot}(k_{\text{target}}).
\label{eq:prefix}
\end{align}

\begin{table*}[thbp]
  \caption{Comparison on FLEURS evaluation set for each language group in CER(\%). }
  \label{table:results4}
  \centering
  \begin{tabular}{l|ccc|ccccccc|c}
    \hline
        ID & \begin{tabular}{c}
joint \\
decoding
\end{tabular}  & \begin{tabular}{c}
decoder \\
prompting
\end{tabular} & \begin{tabular}{c}
encoder \\
prompting
\end{tabular} & WE & EE & CMN & SSA & SA & SEA & CJK & avg. \\
    \hline
    (d) &  $\checkmark$  & & & 14.2 & 11.6 & 12.0 & 16.4 & 18.5 & 18.0 & 19.5 & 15.2  \\
    (h) &  $\checkmark$  &  $\checkmark$  &    & 13.9 & 10.9 & 11.9 & 16.2 & 17.8 & 17.7 & 19.5 & 14.8  \\
    (j) &  $\checkmark$  &  $\checkmark$  &  \textit{Aggregation} &  \bf{13.7} & \bf{10.4} & \bf{11.7} & \bf{16.1} & \bf{15.8} & \bf{16.8} & \bf{19.4} & \bf{14.2} \\
    \hline
  \end{tabular}
\end{table*}

\subsection{Soft prompting}
There may be applications where it is not necessary to recognize all languages, but only a few.
For example, in a service where three languages (e.g. English, Chinese and Japanese) are assumed to be input, it is necessary to perform language identification among the three languages.
The condition that only three languages are input out of more than a hundred language candidates is beneficial for speech recognition.
Our encoder prompting can provide this information to the model.
We propose to extend the encoder prompting approach, \textit{Aggregation} (Eq.(\ref{eq:aggregation})), specifically to multiple languages as:
\begin{align}
\hat{p}_t(k) = \begin{cases}
p_t(k) & k \notin K_{\text{LID}} \\
\frac{\sum_{k' \in K_{\text{LID}}} p_t(k')}{ \sum_{k' \in K_{\text{target}}} p_t(k')} p(k) & k \in K_{\text{target}} \\
0 & \text{otherwise}
\end{cases},
\label{eq:softprompting}
\end{align}
where $K_{\text{target}} \subseteq K_{\text{LID}}$ represents the set of target language candidates.
If the number of candidates is $1$ then this is equivalent to Eq.(\ref{eq:aggregation}).
There is an approach to input language code in multiple languages in prior work, though, but it requires specialized structure for adaptation \cite{9747905}.
They use the one-hot-vectors assigned to each language, normalized to add up to a total of 1.
However, it requires the network to be trained to input language codes.
Our approach can be realized simply by using the general SC-CTC at training time and adapting it using Eq. 4 during inference.


\section{Experimental evaluation}

\subsection{Common Voice + VoxForge}

We evaluated the effectiveness of the proposed method through experiments on various large scale multilingual datasets.
First, we used Common Voice and VoxForge data \cite{ardila2020common,voxforge}.
There were 52 languages in total and 6,000 hours of data in all.
In this paper, we categorized the Common Voice evaluation data into four groups of languages based on the amount of data per language.
High is for languages with more than 100 hours of data, middle is for 20-100 hours, low is for 5-20 hours, and extremely low is for less than 5 hours.
The model sizes used in our experiments were the same except for the intermediate layers.
The encoder was a 12-layer transformer, each with 512 dimensions.
The number of heads of the attention was 4.
The intermediate layer was added at the 6th layer of the encoder, and the weight of the InterCTC was set to 0.3.
The number of tokens in the output was 7000, including 52 language tokens.
Training was further performed by multi-task training of the CTC and the attention decoder \cite{hori2017joint}.
The CTC weights of the multi-task training were also set to 0.3.

Table \ref{table:results1} shows the results.
First, we compared the performance of multilingual models without any prompting ((a) to (d)).
Comparing (a) and (b) confirmed that CTC's joint-decoding was also effective in multilingual speech recognition.
On the other hand, InterCTC (c) and SC-CTC (d) showed a slight performance degradation.
Next, we compared the use of decoder prompting, assuming that the model is being used for a specific language ((e) to (h)).
Decoder prompting improved the performance of the multilingual model, and in all cases, performance was improved over the model without decoder prompting.
Therefore, as mentioned before, it was confirmed that it was benefical to use decoder prompting appropriately when the input language was known.
It was quite interesting to note that when decoder prompting was used as opposed to when not used, the joint decoding degraded the recognition performance.
This provides evidence for our hypothesis that the CTC branch was not appropriately controlled by language IDs.
Finally, we compared the performance of the proposed encoder prompting ((i) to (k)).
\textit{Replacement} (i) and \textit{Aggregation} (j) had almost identical results.
Since the output of CTC tends to be sparse, aggregating probabilities and replacing them with one-hot vectors were almost the same operations.
On the other hand, \textit{Prefix} (k) had almost no improvement over the model without encoder prompting (h).
This was because the original incorrectly estimated language IDs remained in the token sequence by simply modifying the first frame of the utterance.
In particular, encoder prompting with \textit{Replacement} and \textit{Aggregation} was able to significantly improve the performance of low-resource languages with relatively poor baseline performance.

Table \ref{table:results2} shows the language identification accuracy.
There were no significant differences in performance between the encoder types.
However, we observed a tendency for language identification accuracy to decrease with less data.
When the language identification performance was low, the impact of encoder prompting was large because the amount of modification by encoder prompting increased.

Table 3 shows the results of the soft prompting.
In this experiment, as described in Sec. 3.3, we assumed a situation in which three languages (English, Chinese, and Japanese) were input.
In this corpus, English, Chinese, and Japanese are classified as high-, middle-, and low-resource, respectively.
The soft prompting was applied to target these three languages.
We found that soft prompting reduced errors without explicitly providing language information to decoder prompting.


\subsection{FLEURS}

We also evaluated on the FLEURS corpus \cite{conneau2023fleurs}.
FLEURS included 102 languages, with a total of 1.4k hours.
Therefore, on average, it contained a little over 10 hours of data for each language.
In this corpus we also trained a conformer-based SC-CTC.
The conformer encoder was a 12-layer, each with 512 dimensions.
The number of tokens in the output was 6500, including 102 language tokens.
The other settings were same as in Sec. 4.1.
Note that the SSL model was not used since we evaluated the correlation of model performance with the amount of data.
In this experiments, we used \textit{Aggregation} as encoder prompting, which is simpler to implement than \textit{Replacement}.

Table 2 shows the experimental results.
The languages were divided into seven groups as in the previous study \cite{conneau2023fleurs,chen2023improving}.
The improvement using the proposed method was smaller for this corpus than for Common Voice.
This was partly because the domain of the Common Voice data was open domain, while FLEURS was based on Wikipedia.
In this experiment, the average language identification accuracy was $94.2\%$.
This was higher than that of low-resource (5-20 hours) in Table \ref{table:results2}, which had a similar amount of data for each language.
Therefore, the domain difference made language identification relatively easy.
However, we still observed consistent improvements across all language groups, showing the efficacy of our approach.


\section{Conclusion}

In this paper, we proposed a new multilingual speech recognition adaptation technique using encoder prompting.
Encoder prompting allows flexible adaptation during inference in a human-interpretable discrete domain within the encoder.
Our approach, built using SC-CTC, improved recognition performance in situations where the language was predefined.
Also, unlike decoder prompting, soft prompting can be applied to the encoder even in cases where input can be in any one of multiple languages.
We confirmed that recognition performance was also improved in such a setting using soft prompting.
There is still a limitation that encoder prompting must maintain the estimated token sequence length, which is CTC-dependent.
It may be possible to mitigate this constraint by using the attention decoder to output the token sequence and return it to the intermediate layer.
We believe that controlling the encoder as well as the decoder with prompts should receive more attention.

\newpage

\bibliographystyle{IEEEtran}
\input{kashiwagi_encoderprompting.bbl}

\end{document}